# The Separation of Secondary Positrons Produced in the Galaxy from the High Energy Positrons that are Observed by the Recent Space Experiments on PAMELA and AMS-2 and A Comparison of These "Modified" Spectra with Possible Sources of the Excess Positrons Above 10 GeV

W.R. Webber

New Mexico State University, Astronomy Department, Las Cruces, NM  88003, USA

# ABSTRACT

The large intensity of >10 GeV positrons which apparently come from sources outside the Earth-Sun system observed recently by many spacecraft (PAMELA, FERMI, AMS-2) is still a mystery with broad implications. In our attempts to solve this mystery we have first tried to define reasonable limits to the positrons produced in our own galaxy by nuclear interactions of cosmic rays. This is best done by using the secondary B/C ratio produced by these same cosmic rays in order to define the amount of matter traversed by galactic cosmic ray nuclei. Using new values of the B/C ratio together with earlier calculations of positron production by Moskalenko and Strong, 1998, we find that at 10 GeV this galactic production is from 70% to almost 100% of the positrons observed by the above experiments. At 100 GeV these fractions are still from ~20 to 33% of the positrons observed. The resulting "excess" positron spectrum above this normal galactic background is found to have an exponent ~-2.75, possibly flattening at lower energies. If these positrons are coming to us from a uniform source distribution in and beyond the disk of the galaxy with a "source" spectrum that is $\sim E^{-2.0}$, a spectral steepening caused by synchrotron and inverse Compton losses, which are largest near the galactic plane, will produce a spectrum $\sim E^{-2.75}$ at the Earth. This is similar to the "excess" spectrum we find. The excess positron spectrum $\sim E^{-2.75}$ that we obtain is also very similar to the positron spectrum produced by galactic protons in the Earth's atmosphere and in the spacecraft itself which also has a spectrum $\sim E^{-2.75}$. This results in a "background" for all the above experiments. The assumption is that this type of background is removed by the stringent criteria that are imposed on each event. This calculation needs to be reconsidered before the implications of these important positron results can be fully evaluated.

**<u>Introduction</u>**

New positron measurements at energies $\geq$ 10 GeV (Adriani, et al., 2009, 2014; Ackermann, et al., 2012; Aguilar, et al., 2014) on the PAMELA, FERMI and AMS-2 spacecraft have convincingly shown the presence of a high energy component of positrons as part of the high energy electronic component of galactic cosmic rays. This component is in addition to a lower energy secondary positron component produced by galactic cosmic ray protons and heavier nuclei during their propagation through several g/cm$^2$ of interstellar matter. The fraction of these extra higher energy positrons reaches a value ~15% of the total measured electron flux at energies ~300 GeV/nuc (Aguilar, et al., 2014). These AMS-2 measurements do not detect a significant anisotropy ($\leq$ a few %) for these high energy positrons, thus making it more difficult to explain them as coming from a local point source in either the disk and halo of the galaxy.

The positrons from any possible positron source outside of the disk must still pass through the disk en-route to the Earth, however. Thus any spectrum above a few GeV coming from such a source will be steepened by 0.8-1.0 power in the spectral exponent due to synchrotron and inverse Compton energy losses which are large in the disk within $\pm$ 1.0 Kpc of the Sun (Kobayashi, et al., 2004; Webber, 2015).

After first determining a reasonable spectrum for the excess of the observed positrons above a robust galactic secondary positron component, using the latest AMS-2 data as an example; we then use a Monte Carlo diffusion model to calculate the spectra at the Earth from possible sources of positrons located perpendicular to the galactic disk. These propagated spectra are compared with the measured excess of positrons above 10 GeV. The goal of this paper is to try and narrow down the search for the source or sources of these positrons.

There have been many attempts to explain the remarkable AMS-2 positron results at high energies. Our paper has been spurred by the papers by Cowsik, Burch and Madziwa-Nussinov, 2015, Ahlen and Tarle, 2014, and Blum, Katz and Waxman, 2013 among others, which explain the spacecraft observations in terms of features of the galactic production from nuclear interactions of the highest energy protons which, in turn, produce the highest energy positrons. These papers have also been a valuable resource on other previous attempts to explain the spacecraft positron results. These references will not be repeated here. We believe, as is noted by the above authors and others, that the spectral index ~-2.75 that is derived for the excess positrons at high energies is the key to understanding their origin. However, we have two new slants on the significance of this spectral index. These two new aspects of the interpretation will be described in this paper.

**The Data – The Observed Spectrum of Positrons Above a Few GeV**

Here we will only consider positrons above a few GeV and examine the transition between a low energy component of purely galactic "secondary" positrons produced by cosmic ray protons and nuclei in the galaxy and an "additional" positron component that exceeds the secondary component above 10-20 GeV.

First we show in Figure 1 the intensity times $E^3$ of all electrons measured by the PAMELA experiment (Adriani, et al., 2014), the FERMI experiment (Ackermann, et al., 2012) and the AMS-2 experiment (Aguilar, et al., 2014).

In the bottom part of Figure 1 we also show the spectra of positrons measured by PAMELA and by AMS-2 in the same j x $E^3$ representation. These measured positrons spectra are shown along with the original calculated spectrum of secondary positrons (in red) (Moskalenko and Strong, 1998). These secondary galactic positrons are seen to be an important contributor to positrons at all energies. At energies < 10 GeV this galactic secondary positron component is, in fact, the dominant source of all positrons.

It is seen that the two experiments, PAMELA and AMS-2, give roughly similar positron spectra up to ~10 GeV. These spectra lie <u>below</u> the calculated galactic spectrum from Moskalenko and Strong, 1998. This is also the case for the electron intensities as well, where for the galactic spectrum shown in Figure 1 we take the calculations of the LIS electron spectrum by Webber, 2015. The fact that the data for both e+ and e- lies below the predictions is because of the solar modulation of both electrons and positrons below a few GeV. In fact, the fractions that the measured spectra lie below the predicted spectra for both electrons and positrons are quite similar and can be used to delineate any possible positive, negative charge dependence of the overall solar modulation of these components in the heliosphere. The effects of solar modulation on the $e^+ + e^-$ and $e^+$ spectra are shown as shaded regions in Figure 1.

The convergence of the direct measurements of positron intensity and the calculated secondary production by Moskalenko and Strong, 1998, to within about 10-20% of each other at ~10 GeV where modulation effects become small, illustrates the level of accuracy of the secondary propagation calculations for positrons at that energy.

In the j x $E^3$ representation in Figure 1, the calculations of the galactic intensity of positrons decreases above 10 GeV indicating a spectrum that is steeper than $E^{-3.0}$. This is after propagation in the disk of the galaxy which steepens the source spectrum. The source spectrum for these positrons is ~$E^{-2.75}$ as noted by Moskalenko and Strong, 1998, thus the positron production spectrum itself is notably flatter than $E^{-3.0}$.

Both the PAMELA and AMS-2 intensities of total positrons slowly <u>increase</u> above 10 GeV in the j x $E^3$ representation which means that the spectrum is flatter than $E^{-3.0}$, which is more like a production spectrum from nuclei interactions. Note that at ~100 GeV, the normal galactic secondary production from nuclei is still ~20% of the AMS-2 measurements at that energy according to the calculations of Moskalenko and Strong, 1998.

The calculations of Moskalenko and Strong, 1998, are based on a propagation path length which in turn is based on the measured B/C ratio. At 100 GeV/nuc their model predicts a B/C ratio of 0.070 and at 200 MeV/nuc this ratio is predicted to be 0.050. These values are based on a cosmic ray escape length which is taken to be ~$P^{-0.60}$ at energies above a few GeV. Recent measurements of the B/C ratio by AMS-2 at the energies of 100 and 200 GeV/nuc are 0.115 and 0.095 respectively (Oliva, et al., 2015). This indicates that the production of positrons at the corresponding energies originally calculated by Moskalenko and Strong, 1998, should be increased by factors of about 1.65 and 1.90 respectively. At 100 GeV this enhanced galactic secondary positron production would now be ~33% of the total positrons observed by AMS-2.

In Figure 2 we show the "excess" positron spectrum, defined here as the ratio of the AMS-2 measurements to the production spectrum of Moskalenko and Strong, 1998, shown as two separate curves, one corresponding to the original prediction (low), the other fitting the Moskalenko and Strong calculations to the new AMS-2 measurements of the B/C ratio (high) at 100 and 200 GeV/nuc.

In Figure 3 we show these "low" and "high" "excess" spectra as j vs. $E^3$ spectra. The individual "data" points are an average $(j_{Hi} - j_{Lo})/2$ at each energy. At energies up to ~100 GeV the very uncertain spectral index of this excess component appears to be increasing from about -2.0 at 30 GeV to -2.75 at 100 GeV. Above 100 GeV, where this index can be more accurately determined, the index remains at approximately ~2.75 up to the highest energy point at ~300 GeV. This is the spectrum of positrons that is unaccounted for by the galactic secondary production and needs to be explained. Note that this spectral index of "excess" positions is very similar to the spectrum of primary galactic protons also measured simultaneously by AMS-2. This proton spectrum has a slope between -2.7-2.8 above ~30 GeV (e.g., Choutko, et al., 2015).

This spectral similarity between the two cosmic ray species, protons and positrons, is striking and needs to be understood since these protons also create most of the "background" in all of the above spacecraft measurements in Earth orbit which can turn out to be a very complex temporal and angular calculation including also the Earth's atmosphere as well as the spacecraft itself..

**Calculations Related to a Positron Source $\perp$ to the Galactic Plane**

As a further step in trying to understand the origin of this excess in positrons we have calculated how the spectrum from a possible positron source in a direction $\perp$ to the galactic plane would be modified by diffusive propagation through the galactic magnetic field to the Earth assumed to be located on the galactic plane.

This calculation uses a Monte Carlo Diffusion Model (MCDM) (see Webber, 2015, and earlier references there-in) to determine the fraction of particles emitted from a source at distances between Z=0 and Z=1.0 Kpc perpendicular to the galactic plane that will arrive at the Sun. The calculation is similar to the calculation of the galactic cosmic ray electron distributions at various energies perpendicular to the plane of the galaxy as shown in Figure 5 of Webber, 2015. The magnetic field is assumed to be =5μG at Z=0 and to decrease exponentially with an e-folding distance = 1.5 Kpc. The diffusion coefficient is assumed to be equal to 2 x $10^{28}$ cm$^2$·s at 1 GeV at Z=0 and to be ~$E^{0.5}$ above 1 GeV (see also Webber and Higbie, 2008, for other parameters).

7We assume a positron "source" spectrum which is $\sim E^{-2}$, as might possibly occur in dark matter models. The spectrum that would be observed at the Sun from a uniform distribution of these hypothetical sources between zero and 1.0 Kpc from the plane is shown in Figure 3. Superimposed on this spectrum are the average high and low excess spectra deduced from the subtraction of the galactic production from the AMS-2 measurements. This excess AMS-2 positron spectrum is normalized to have the same intensity as the Monte Carlo calculations at 100 GeV. For sources beyond the boundary, here taken to be =1.5 Kpc, the positron source intensities will decrease dramatically, but retain the same spectral shape.

**Summary and Discussion**

The excess positron spectra obtained from the AMS-2 measurements reported by Aguilar, 2014, and which are obtained in this paper are shown in Figures 2 and 3. These spectra can be reproduced in at least two ways and the "source" spectra for this excess is not yet uniquely determined.

First of all we do note that this excess spectrum of positrons has essentially the same spectral index of -2.75 as the observed spectra of protons and other galactic nuclei. These particles would be the input to the determination of the instrumental "positron" backgrounds that would be present in each observation. At present these backgrounds are believed to be fully included in each presented spectra. However, this coincidence of spectral indices between the excess spectrum and a possible instrumented background spectrum makes it imperative that these background calculations be re-examined for all measurements.

Secondly, the Monte Carlo calculations that we have done show that the excess spectra of positrons on AMS-2 as determined in this paper, which has a spectral slope ~2.75 above 100 GeV possibly flattening at lower energies, could be reproduced reasonably well by a uniform source distribution of an unknown component of positrons perpendicular to the plane of the galaxy, which itself has a spectrum $\sim E^{-2.0}$. The synchrotron and inverse Compton energy losses near the plane of the galaxy will steepen the positron source spectrum to one which is $\sim E^{-2.75}$, similar to the "excess" spectrum obtained from the AMS-2 measurements.

**Acknowledgments:** The author appreciates discussions with Paul Higbie. This article could not have been written without the assistance of Tina Villa.

## Figure Captions

**Figure 1:** The energy spectra of cosmic ray ($e^+ + e^-$) and $e^+$ in a j x $E^3$ format. The total e intensities are from PAMELA (Adriani, et al., 2014), FERMI (Ackermann, et al., 2012) and AMS-2 (Aguilar, et al., 2014). The $e^+$ data is from PAMELA (Adriani, et al., 2009) and AMS-2. Also shown is the galactic production of $e^+$ calculated by Moskalenko and Strong, 1998. Note that the calculated and observed $e^+$ intensities cross at about 10 GeV, indicating the excess positron intensity would be zero at this energy. Also shown is the calculated galactic e- spectrum by Webber, 2015, obtained using a source spectrum with a single spectral index ~-2.25, normalized to the AMS-2 intensity measured at 10 GeV plus 10% (to allow for solar modulation).

**Figure 2:** Measured $e^+$ spectrum by AMS-2 in j x $E^3$ format (solid line with points) along with two possible limits of galactic $e^+$ production (in red). LO is from the original Moskalenko and Strong, 1998 calculation. HI is the original calculation of both the e+ intensity and the B/C ratio made by Moskalenko and Strong, 1998, times the B/C ratio actually measured by AMS-2 (Oliva, et al., 2015) at 100 and 200 GeV/nuc, to these original calculations of this ratio. The corresponding limits on the excess positron spectrum from the AMS-2 measurements with the red curves subtracted as secondary production are shown in purple for the LO and HI production calculations.

**Figure 3:** The AMS-2 excess positron spectra shown in a j x $E^3$ representation. The average excess spectrum used here is defined as $(j_{Hi} + j_{Lo})/2$ and is shown as a black solid line. The "error" bars represent limits from the values of the HI and LO excess spectra. Note that this spectral index is ~-2.75 above 100 GeV where it is best defined. Below 100 GeV the spectral index becomes smaller (and the uncertainties get larger). We recognize that the excess intensity should go to zero at about 10 GeV. Also shown in this figure is a spectrum obtained from a uniform source distribution perpendicular to the plane of the galaxy with a source exponent = $E^{-2.0}$.

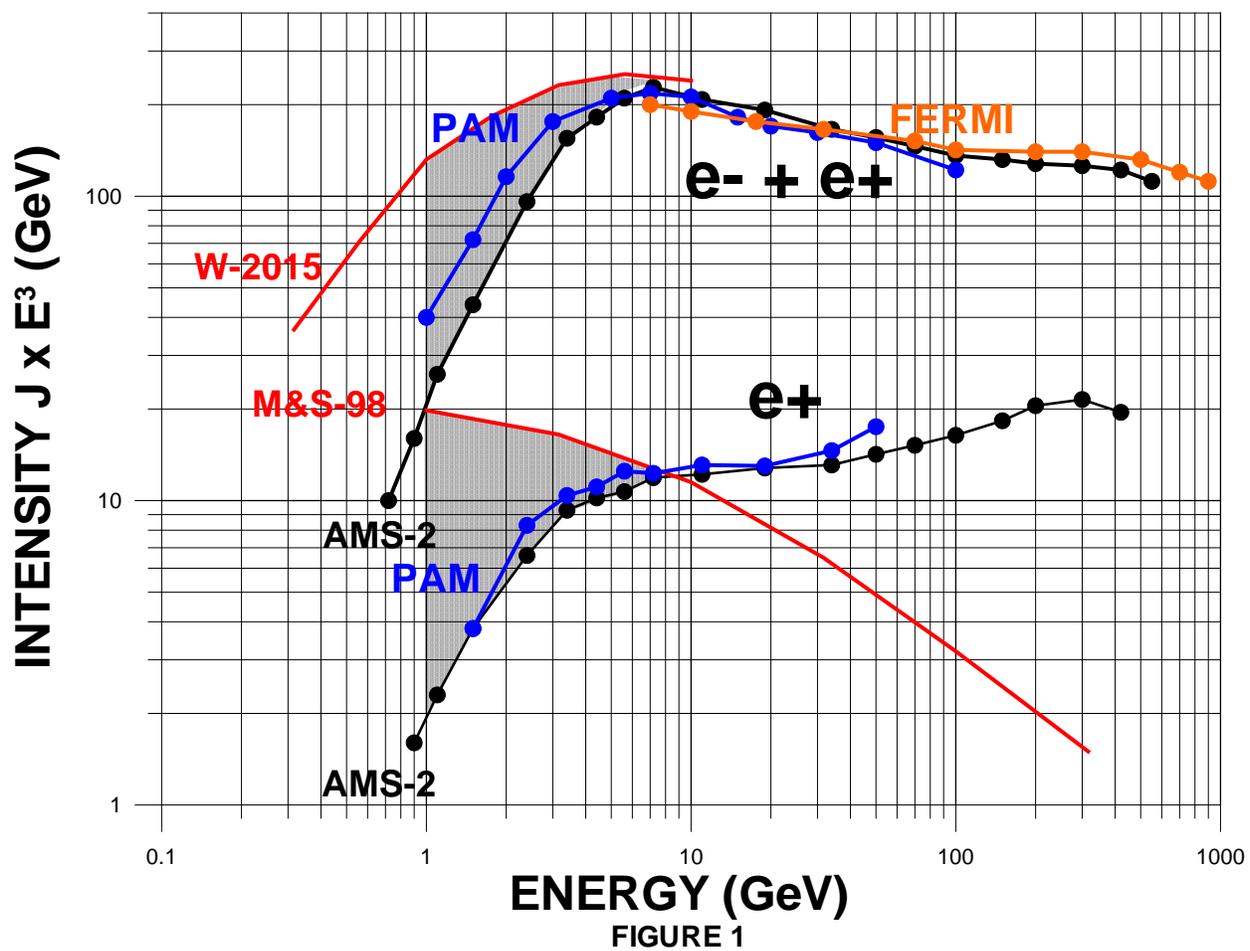

**FIGURE 1**

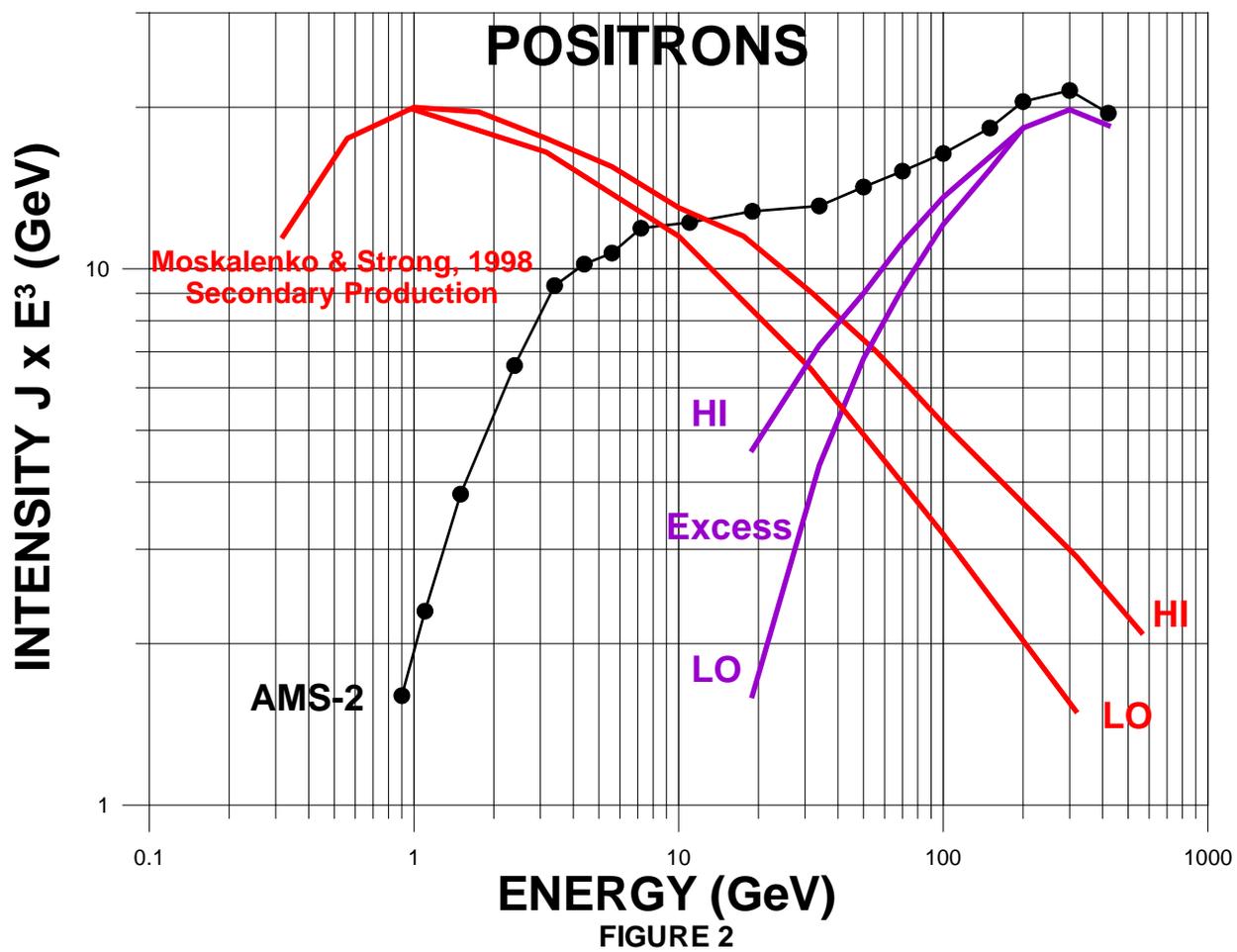

FIGURE 2

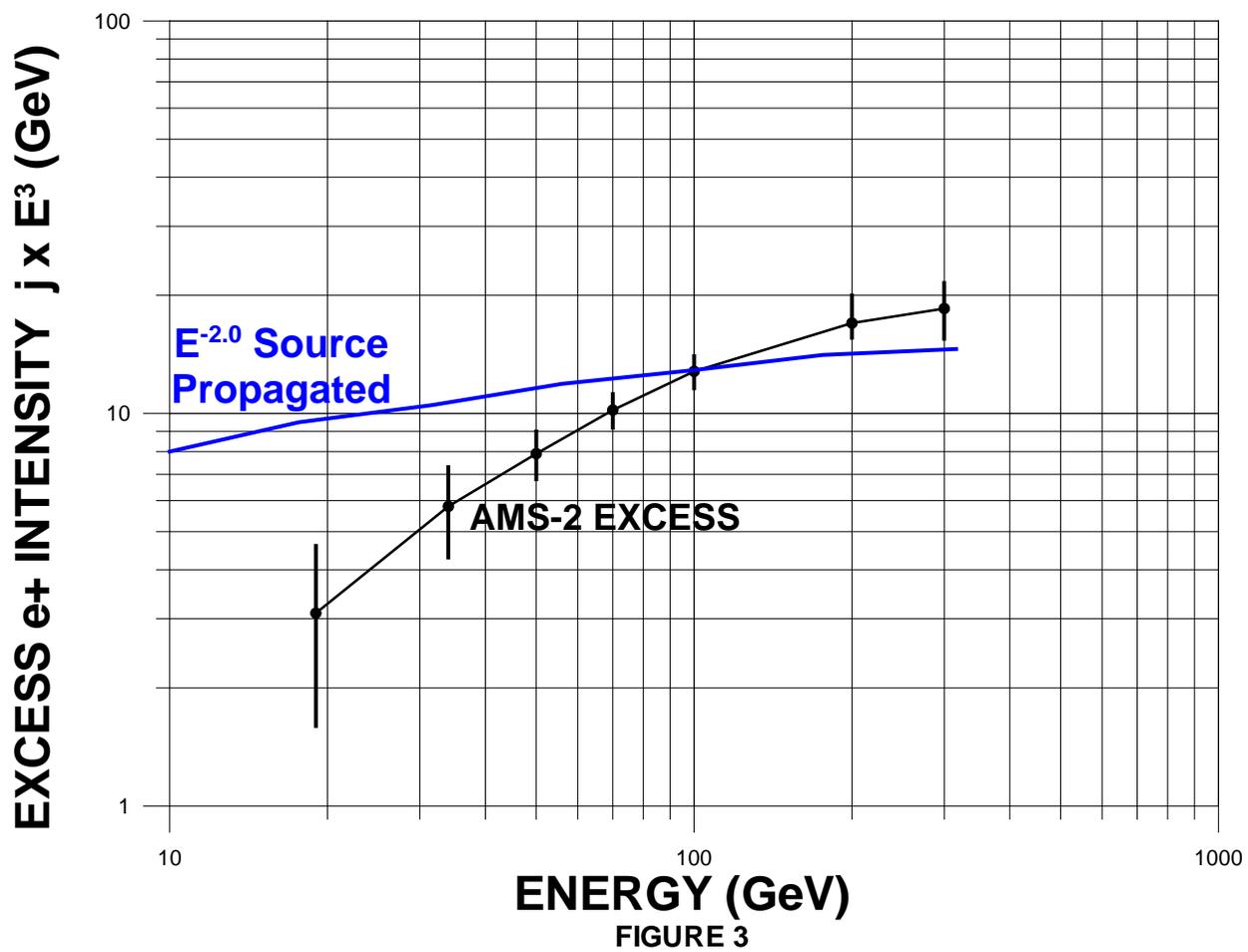

FIGURE 3